\documentclass[sigconf]{acmart}

\usepackage{array}
\usepackage{balance}
\usepackage{subfigure}
\usepackage{multirow}
\usepackage{float}
\usepackage{soul}
\usepackage{mdframed}
\usepackage{amsopn}
\usepackage{mathrsfs}
\usepackage{mathtools}
\usepackage{amsmath}
\usepackage{bbm}
\usepackage{arydshln}
\usepackage{hyperref}
\usepackage{multicol}
\usepackage{blkarray}
\usepackage{enumitem}
\usepackage{courier}
\usepackage{rotating}
\usepackage{booktabs}
\usepackage{diagbox}
\usepackage{fancybox}
\usepackage{minibox}
\usepackage{cases}
\usepackage[noend]{algpseudocode}
\usepackage{algorithm}
\usepackage{algorithmicx}
\usepackage{xcolor}
\usepackage{adjustbox}

\definecolor{littlestrawberry}{RGB}{255,102,178}
\definecolor{littlemango}{RGB}{255,153,51}

\newcommand{\argmax}{\operatornamewithlimits{argmax}}

\newtheorem{definition}{Definition}

\newcommand{\approach}{\textrm{\textsc{Cafe}}}

\AtBeginDocument{%
  \providecommand\BibTeX{{%
    \normalfont B\kern-0.5em{\scshape i\kern-0.25em b}\kern-0.8em\TeX}}}

\copyrightyear{2020}
\acmYear{2020}
\setcopyright{acmcopyright}\acmConference[CIKM '20]{Proceedings of the 29th ACM International Conference on Information and Knowledge Management}{October 19--23, 2020}{Virtual Event, Ireland}
\acmBooktitle{Proceedings of the 29th ACM International Conference on Information and Knowledge Management (CIKM '20), October 19--23, 2020, Virtual Event, Ireland}
\acmPrice{15.00}
\acmDOI{10.1145/3340531.3412038}
\acmISBN{978-1-4503-6859-9/20/10}

\begin{document}
\fancyhead{}
\title[\textsc{Cafe}: Coarse-to-Fine Neural Symbolic Reasoning for Explainable Recommendation]{\textsc{Cafe}: Coarse-to-Fine Neural Symbolic Reasoning\\for Explainable Recommendation}

\author[Xian et al.]{Yikun Xian$^{\dagger}$, Zuohui Fu$^{\dagger}$, Handong Zhao${^\ddagger}$, Yingqiang Ge${^\dagger}$, Xu Chen${^\S}$, Qiaoying Huang${^\dagger}$,\quad\quad\quad Shijie Geng${^\dagger}$, Zhou Qin${^\dagger}$, Gerard de Melo$^\mathparagraph$, S.\  Muthukrishnan${^\dagger}$, Yongfeng Zhang${^\dagger}$}
\renewcommand{\authors}{Yikun Xian, Zuohui Fu, Handong Zhao, Yingqiang Ge, Xu Chen, Qiaoying Huang, Shijie Geng, Zhou Qin, Gerard de Melo, S.\  Muthukrishnan, Yongfeng Zhang} %
\affiliation{%
  \institution{$^{\dagger}$Rutgers University, NJ \quad $^{\ddagger}$Adobe Research, CA \quad $^\S$University College London, UK \quad $^\mathparagraph$HPI/Univ.\ of Potsdam, Germany}
}
\email{siriusxyk@gmail.com, zuohui.fu@rutgers.edu, 	hazhao@adobe.com, yingqiang.ge@rutgers.edu, xu.chen@ ucl.ac.uk}
\email{{qh55, sg1309, zq58, gerard.demelo}@rutgers.edu,  muthu@cs.rutgers.edu, yongfeng.zhang@rutgers.edu}

\renewcommand{\shortauthors}{}

\begin{abstract}
Recent research explores incorporating knowledge graphs (KG) into e-commerce recommender systems, not only to achieve better recommendation performance, but more importantly to generate explanations of why particular decisions are made. This can be achieved by explicit KG reasoning, where a model starts from a user node, sequentially determines the next step, and walks towards an item node of potential interest to the user. However, this is challenging due to the huge search space, unknown destination, and sparse signals over the KG, so informative and effective guidance is needed to achieve a satisfactory recommendation quality. 
To this end, we propose a CoArse-to-FinE neural symbolic reasoning approach (\approach{}). It first generates user profiles as coarse sketches of user behaviors, which subsequently guide a path-finding process to derive reasoning paths for recommendations as fine-grained predictions. User profiles can capture prominent user behaviors from the history, and provide valuable signals about which kinds of path patterns are more likely to lead to potential items of interest for the user. To better exploit the user profiles, an improved path-finding algorithm called Profile-guided Path Reasoning (PPR) is also developed, which leverages an inventory of neural symbolic reasoning modules to effectively and efficiently find a batch of paths over a large-scale KG. We extensively experiment on four real-world benchmarks and observe substantial gains in the recommendation performance compared with state-of-the-art methods.

\end{abstract}


\maketitle

\section{Introduction}\label{sec:intro}
\begin{figure}
\centering
\includegraphics[width=\linewidth]{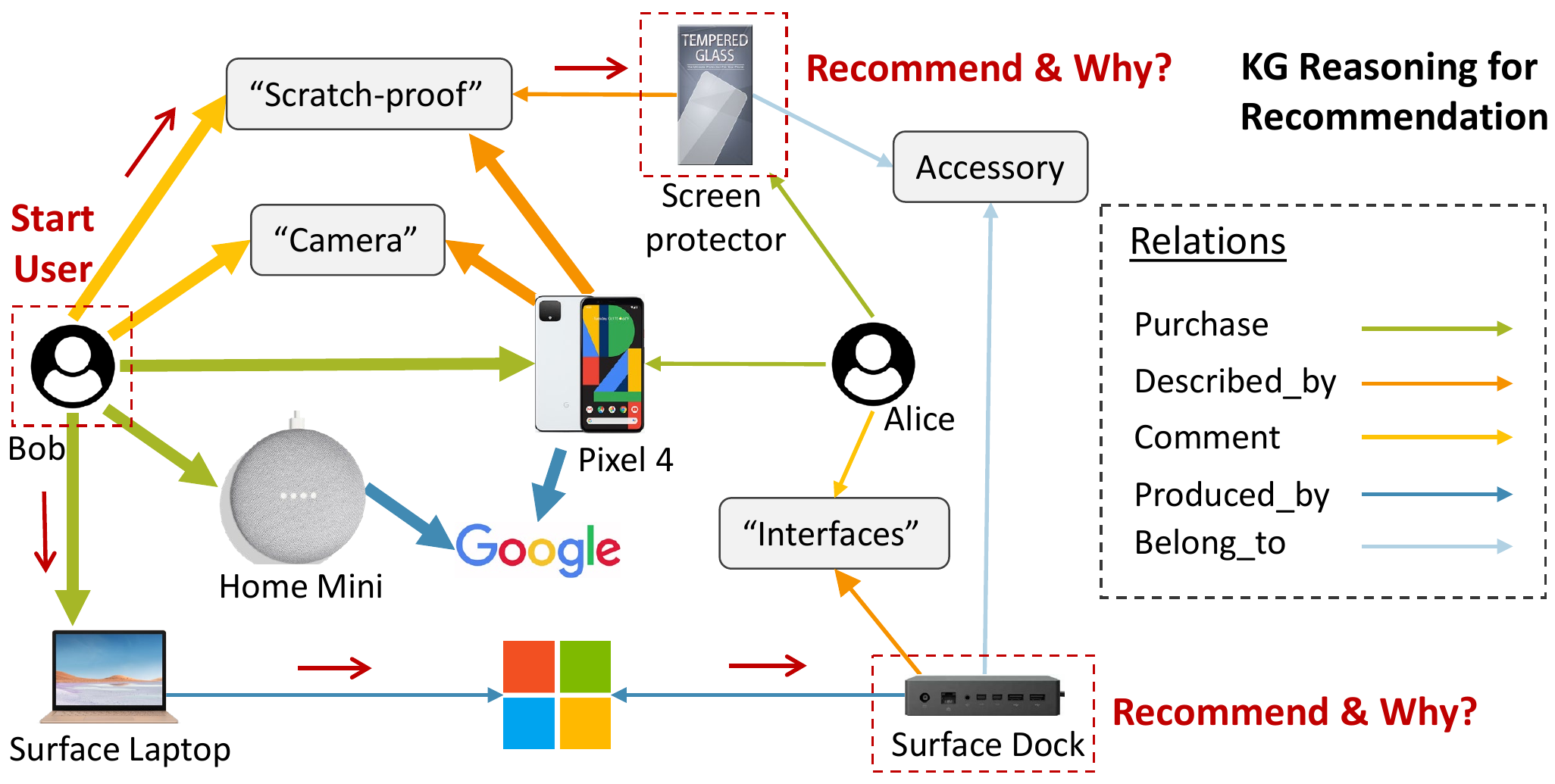}
\vspace{-15pt}
\caption{A motivating example of KG reasoning for e-commerce recommendation. Given the start user, the target destinations (i.e., items to recommend) are unknown beforehand. 
The goal is -- guided by user behavior patterns (bold edges) -- to sequentially determine the next step traversing the KG towards  potential items of interest as recommendations (e.g., \textit{\textbf{Screen protector}} and \textit{\textbf{Surface Dock}}). Two possible reasoning paths are marked with red arrows, which are taken as explanations to the recommendations.}
\label{fig:motivation}
\end{figure}

Recommender systems on modern e-commerce platforms serve to support the personalization of the customer shopping experience by presenting  potential products of interest to users \cite{schafer2001commerce,ge2020understanding}.
They draw on diverse forms of historical user behavior, including but not limited to past browsing and previously purchased products, written reviews, as well as added favorites \cite{Dong2020InterpretableRecommendation}.
The models are expected to capture customized patterns of user preference across products, and hence can be leveraged to provide more accurate recommendations \cite{DBLP:conf/ijcai/0001Z20}.
In addition to accuracy-driven recommendation, it has become increasingly important in modern e-commerce systems to present auxiliary explanations of the recommendations \cite{zhang2018explainable}, i.e., the system aims to supply customers with product recommendations accompanied by informative explanations about why those products are being recommended.

In this regard, knowledge graphs (KG) \cite{KnowledgeGraphs2020} have recently come to prominence to address both requirements.
A KG can not only provide abundant information about users and items, but can also enable explainable recommendations via explicit KG reasoning 
\cite{ai2018learning,xian2019kgrl,wang2019knowledge}: 
Starting from a user node, the system sequentially determines the next-hop nodes, and moves towards potential items of interest for the user.
The derived path explicitly traces the decision-making process and can naturally be regarded as an explanation for the recommended item.
For instance, as shown in Fig.~\ref{fig:motivation}, one possible reasoning path is $\text{User}\xrightarrow{\text{Comment}}\text{``Scratch-proof''}\xrightarrow{\text{Described\_by}^{-1}}\text{``Screen protector''}$, where the product ``Screen protector'' is directly used as a recommendation.

Although KG reasoning for explainable recommendation is promising, several issues still remain to be addressed.
First, 
in order to make use of the reasoning paths to explain the decision-making process, the recommendations are supposed to be derived along with the KG reasoning.
However, many existing approaches \cite{ai2018learning,wang2019kgat} first predict the items to be recommended, and subsequently conduct a separate search for paths matching the user--item pair.
Addressing these tasks in isolation means that the explanation may not reflect the actual decision making process for the recommendation.
Moreover, this fails to allow the recommendation decision making to benefit from the KG reasoning process.
We discuss this further in Section \ref{sec:paradigm}.

Second, previous work on KG reasoning has largely neglected the diversity of user behavior in the historical activity data.
Most approaches consider only item-side knowledge integrated from external sources, such as Freebase \cite{zhao2019kb4rec,wang2019knowledge} or product graphs \cite{ai2018learning,dong2018challenges}, restricting user-side information to simple user interactions (e.g., purchasing a product or rating a movie).
However, in e-commerce recommendation, user purchases may be triggered by different aspects of past behavior.
As an example, in Fig.~\ref{fig:motivation}, the user having purchased product ``Pixel 4'' may contribute to the keyword ``Camera'' that the user mentioned in the comment, or to the brand (``Google'') of some product (``Home Mini'') owned by the user.
User behavior patterns of this sort can be extracted to guide future recommendations (``Screen protectors'' or ``Surface Dock'').

Last but not least, a lack of effective guidance on path reasoning makes it less efficient in finding potential paths in the large search space of the KG.
Due to the large scale of the KG and the unknown destination before path-finding, in practice, it is infeasible to follow previous methods that enumerate paths among all user--item pairs to choose the best one.
Other works \cite{xian2019kgrl,lin2018multi} adopt reward shaping from reinforcement learning \cite{sutton2018reinforcement} to alleviate the issue. However, the reward signal is sparse and cannot effectively and efficiently guide the model to arrive at correct items for recommendation.

In this paper, 
we seek to answer the following three questions regarding the task of KG reasoning for explainable recommendation:
1) Instead of isolating recommendation and path-finding, how to directly perform path reasoning to arrive at items of interest so that the derived paths can explain the recommendation process?
2) Besides rich item-side information, how to explicitly model diverse user behaviors from historical activities so that they can be exploited to provide good guidance in finding potential paths?
3) Upon modeling behavior, how to exploit the user model to conduct the path reasoning in a both effective and efficient manner?

To this end, we propose a CoArse-to-FinE neural symbolic reasoning method (\approach{}), which first generates a coarse sketch of past user behavior, and then conducts path reasoning to derive recommendations based on the user model for fine-grained modeling.
We draw inspiration from the literature in linguistics \cite{pullum2010land,bateman2003natural}, where the human writing process consists of multiple stages focusing on different levels of granularity.
This has also been invoked in NLP tasks such as long review generation, where coarse-level aspects are first sketched to guide the subsequent long text generation \cite{li2019generating,dong2018coarse,fu2020absent}.
In this work, we first compose a personalized user profile consisting of diverse user-centric patterns, each of which captures prominent coarse-grained behavior from historical user activities.
Each profile can provide effective guidance on what patterns of reasoning paths may more likely lead to potential items of interest for a given user.
To fully exploit the profile, we maintain an inventory of neural symbolic reasoning modules and accordingly design a path-finding algorithm to efficiently conduct batch path reasoning under the guidance of such profiles.
Recommendations are consequently acquired from the batch of reasoning paths produced by the algorithm.

This paper makes four key contributions.
\begin{itemize}[leftmargin=*]
\item First, we highlight important shortcomings of past KG reasoning approaches for explainable recommendation, where path-reasoning and recommendation are addressed in isolation.
\item Second, we introduce a coarse-to-fine paradigm to approach the problem by explicitly injecting diverse user behavior modeling into the KG reasoning process.
\item Third, we propose a novel profile-guided path reasoning algorithm with neural symbolic reasoning modules to effectively and efficiently find potential paths for recommendations.
\item Fourth, we experiment on four real-world e-commerce datasets showing that our model yields high-quality recommendation results and the designed components are effective. 
\end{itemize}

\section{Preliminaries}
\subsection{Concepts and Notations}\label{sec:concepts}
In e-commerce recommendation, a \emph{knowledge graph} (or \emph{product graph}) denoted by $\mathcal{G}_\mathrm{p}$ is constructed to capture rich meta-information of products on the platform.
It is defined to be a set of triples, $\mathcal{G}_\mathrm{p}=\{(e,r,e')\mid e,e'\in\mathcal{E}_\mathrm{p},r\in\mathcal{R}_\mathrm{p}\}$, where $\mathcal{E}_\mathrm{p}$ and $\mathcal{R}_\mathrm{p}$ respectively denote the sets of entities and relations.
A special subset of entities are called products (items), denoted by $\mathcal{I}\subseteq\mathcal{E}_\mathrm{p}$.
Each triple $(e,r,e')\in\mathcal{G}_\mathrm{p}$ represents a fact indicating that head entity $e$ interacts with tail entity $e'$ through relation $r$.

At the same time, diverse user activities can also be modeled as a heterogeneous graph denoted by $\mathcal{G}_\mathrm{u}=\{(e,r,e')\mid e,e'\in\mathcal{E}_\mathrm{u},r\in\mathcal{R}_\mathrm{u}\}$, where $\mathcal{E}_\mathrm{u}$ and $\mathcal{R}_\mathrm{u}$ are entity and relation sets satisfying that user set $\mathcal{U}\subseteq\mathcal{E}_\mathrm{u}$, item set $\mathcal{I}\subseteq\mathcal{E}_\mathrm{u}$, and user--item interaction $r_{ui}\in\mathcal{R}_\mathrm{u}$.
When $|\mathcal{R}_\mathrm{u}|=1$ and $\mathcal{E}_\mathrm{u}=\mathcal{U}\cup\mathcal{I}$, $\mathcal{G}_\mathrm{u}$ is a bipartite user--item graph.
Here, we assume $\mathcal{G}_\mathrm{u}$ is the general user interaction graph consisting of diverse interactions and objects, e.g., a user can make comments as in Fig.~\ref{fig:motivation}.

For convenience, we unify both product graph and user interaction graph into the same framework, which we call \emph{User-centric KG}, denoted as $\mathcal{G}=\mathcal{G}_\mathrm{p}\cup\mathcal{G}_\mathrm{u}$ with combined entity set $\mathcal{E}=\mathcal{E}_\mathrm{p}\cup\mathcal{E}_\mathrm{u}$ and relation set $\mathcal{R}=\mathcal{R}_\mathrm{p}\cup\mathcal{R}_\mathrm{u}$.
In the remainder of this paper, the term KG generally refers to this User-centric KG.

A \emph{path} in the KG is defined as a sequence of entities and relations, denoted by $L=\{e_0,r_1,e_2,\ldots,r_{|L|},e_{|L|}\}$ (or simply $L_{e_0\leadsto e_{|L|}}$),
where $e_0,\ldots,e_{|L|}\in\mathcal{E}$, $r_1,\ldots,r_{|L|}\in\mathcal{R}$ and $(e_{t-1},r_t,e_t)\in\mathcal{G}$ for $t=1,\ldots,|L|$.
To guarantee the existence of paths, inverse edges are added into the KG, i.e., if $(e,r,e')\in\mathcal{G}$, then $(e', r^{-1}, e)\in\mathcal{G}$, where $r^{-1}$ denotes the inverse relation with respect to $r\in\mathcal{R}$.
One kind of path of particular interest is called a \emph{user-centric path}. Such a path originates at a user entity ($e_0\in\mathcal{U}$) and ends with an item entity ($e_{|L|}\in\mathcal{I}$).
We also define a \emph{user-centric pattern} $\pi$ to be a relational path between a user and an item, $\pi=\{r_1,\ldots,r_{|\pi|}\}$.
Hence, the relation sequence of any user-centric path forms a user-centric pattern.
Such a pattern can be viewed as a semantic rule that describes a specific user behavior towards a product via some actions (relations) on the e-commerce platform.
Additionally, we define the \emph{user profile} $\mathcal{T}_u$ of user $u$ to be an aggregation of user-centric patterns with weights, $\mathcal{T}_u=\{(\pi_1,w_1),\ldots,(\pi_{|\mathcal{T}_u|},w_{|\mathcal{T}_u|})\}$, where $w_1,\ldots,w_{|\mathcal{T}_u|}\in \mathbb{N}$ are the weights of patterns.
Each user profile distinctively characterizes prominent user behavior from the purchase history as well as diverse other activities, and can be leveraged to guide KG reasoning for recommendation (Section \ref{sec:fine}).

\subsection{Problem Formulation}
In this work, we study the problem of KG reasoning for explainable recommendation in an e-commerce scenario \cite{xian2019kgrl}.
By leveraging rich information in the KG, we aim to predict a set of items as recommendations for each user along with the corresponding user-centric paths as the explanation.
The problem is formulated as follows.
\begin{definition}{(Problem Definition)}\label{def:problem}
Given an incomplete user-centric KG $\mathcal{G}$ and an integer $K$,  for each user $u\in\mathcal{U}$, the goal is to generate 1) a set of $K$ items $\left\{i^{(k)} \big| i^{(k)}\in\mathcal{I},(u,r_{ui},i^{(k)})\not\in\mathcal{G},k\in[K]\right\}$, and 2) $K$ corresponding user-centric paths $\left\{L_{u\leadsto i^{(k)}}\right\}_{k\in[K]}$.
\end{definition}

\subsection{A Coarse-to-Fine Paradigm}\label{sec:paradigm}
The general framework to approach the problem in Def.~\ref{def:problem} consists of two parts: a recommendation component $f_\mathrm{rec}$ and a path inference component $f_\mathrm{path}$.
In most existing approaches \cite{ai2018learning,wang2018ripplenet,wang2019kgat,ma2019jointly,xian2019kgrl}, $f_\mathrm{rec}:u,i\mapsto \mathbb{R}$ estimates a similarity score between user $u$ and an item $i$ using enriched information from the KG. $f_\mathrm{path}:u,i\mapsto L$ outputs a user-centric path $L$ given user $u$ and item $i$ (sometimes $i$ is not necessary as input \cite{xian2019kgrl}).
The major differences between existing works lie in 1) the technical implementation and 2) the composition and execution order of these components. Below we revisit the existing KG reasoning paradigms and highlight the benefits of the proposed coarse-to-fine paradigm.

\paragraph{Rec-First Paradigm}
One group of approaches \cite{ai2018learning,wang2018ripplenet,wang2019kgat} first makes recommendations via $f_\mathrm{rec}$, followed by a separate process $f_\mathrm{path}$ to search paths that best match the predicted user--item pair:
\begin{align*}
\hat{i} = \argmax_{i\in\mathcal{I}} f_\mathrm{rec}(u,i;\mathcal{G}),\quad
\hat{L}_{u\leadsto\hat{i}} = f_\mathrm{path}(u,\hat{i};\mathcal{G}),
\end{align*}
where $\hat{i},\hat{L}_{u\leadsto\hat{i}}$ are the predicted item and path, respectively.
Common choices of $f_\mathrm{rec}$ include KG embeddings \cite{bordes2013translating,wang2018ripplenet} and relational graph neural networks \cite{schlichtkrull2018modeling,wang2019kgat}.
$f_\mathrm{path}$ usually refers to a path ranking model \cite{xian2019kgrl,fu2020fairness} or graph search algorithm \cite{cormen2009introduction,ai2018learning}.
However, it is worth noting that one critical limitation of this paradigm is the isolation of recommendation $f_\mathrm{rec}$ and path selection $f_\mathrm{path}$.
This may degrade recommendation performance, as it is solely determined by $f_\mathrm{rec}$, but fails to benefit from the post-hoc path-finding of $f_\mathrm{path}$. More importantly, the reported path is not a genuine explanation of the actual recommendation process.

\paragraph{Path-Guided Paradigm}
Another line of work \cite{xian2019kgrl} first uses $f_\mathrm{path}$ to perform path-finding with unknown destination and the reached item is naturally adopted as the recommendation:
\begin{align*}
\hat{L}_{u\leadsto e_{T}} = f_\mathrm{path}(u,-;\mathcal{G}_\mathrm{u}),\quad
\hat{i} = e_T,
\end{align*}
where ``$-$'' means no item is required as input and $e_{T}$ is the last entity of path $\hat{L}_{u\leadsto e_{T}}$.
Here, $f_\mathrm{path}$ usually adopts a multi-step reasoning model such as a policy network \cite{sutton2018reinforcement,lin2018multi,xian2019kgrl} to sequentially pick the next step in the KG.
Since the recommendation is derived along with the path inference results, $f_\mathrm{path}$ implicitly contains the recommendation process, and the resulting path can be used to track and explain the decision-making process.
However, due to the challenges of unknown destinations and the huge search space of KG, the signals (e.g., rewards) are very sparse and cannot effectively guide the path inference to achieve satisfactory recommendation, in comparison with Rec-First approaches.

\paragraph{Coarse-to-Fine Paradigm}
To achieve direct path reasoning while simultaneously obtaining  competitive recommendation performance, we propose a novel coarse-to-fine paradigm.
In the coarse stage, we introduce a new component $f_\mathrm{profile}:u\mapsto \mathcal{T}_u$ that composes a user profile $\mathcal{T}_u$ to capture prominent user behavior from historic data (details in Section \ref{sec:coarse}).
Then, for fine-grained modeling, an improved variant of path inference component $f_\mathrm{path}':u,\mathcal{T}_u\mapsto L$ is developed to perform multi-step path reasoning guided by the composed user profile (details in Section \ref{sec:fine}):
\begin{align}
\mathcal{T}_u = f_\mathrm{profile}(u;\mathcal{G}_\mathrm{u}),\quad
\hat{L}_{u\leadsto e_T} = f_\mathrm{path}'(u,\mathcal{T}_u;\mathcal{G}_\mathrm{u}),\quad
\hat{i} = e_T.
\end{align}
The path reasoning relies on a one-step reasoner $\phi$ with learnable parameter $\Theta$ (see Section \ref{sec:module}).
It determines the $t^\mathrm{th}$ step action by estimating the probability $P_\Theta(r_t,e_t|u,h_t)$ of choosing an outgoing edge $(r_t,e_t)$ given user $u$ and history trajectory $h_t=\{r_1,e_1,\ldots,r_{t-1},e_{t-1}\}$.
Therefore, we can estimate the probability of a multi-step path $L=\{u,r_1,e_1,\ldots,r_T,e_T\}$ being generated by $\phi$:
\begin{align}\label{eq:prob}
\log P_\Theta(L|u) = \sum_{t=1}^{T}\log P_\Theta(r_t,e_t|u,h_t)
\end{align}

\begin{figure*}[t]
    \centering
    \includegraphics[width=\textwidth]{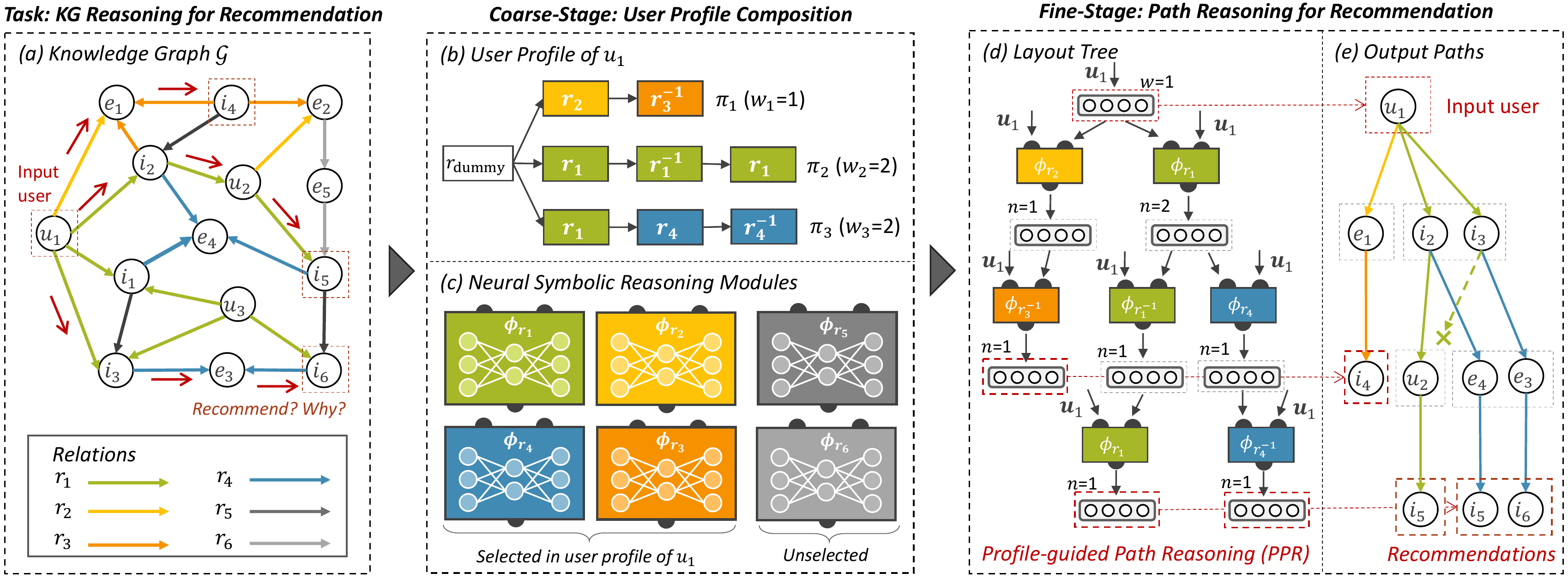}
    \vspace{-20pt}
    \caption{Illustration of \approach{}, a coarse-to-fine KG reasoning approach. (a) Given a KG and a start user, the goal is to conduct multi-step path reasoning to derive recommendations. (b) In the coarse stage, a personalized user profile is constructed based on historic user behavior in the KG. (c) To make use of the user profile in path reasoning, an inventory of neural symbolic reasoning modules is maintained. (d) In the fine stage, a layout tree is composed with the modules based on the user profile, which is exploited by the proposed PPR algorithm (Alg.~\ref{alg:program-exe}) to produce (e) a batch of paths along with recommendations.
    }
    \label{fig:method}
\end{figure*}

\noindent This paradigm has three notable benefits.
\begin{itemize}[leftmargin=*]
\item Explicit user modeling from $f_\mathrm{profile}$ can detect prominent user-centric patterns, which assist the path reasoning process in arriving at potential items of interest to the user.
\item Path inference via $f_\mathrm{path}'$ is conducted under the guidance of the user profile so as to improve both the effectiveness and efficiency of the path-finding process.
\item %
The reasoner $\phi$ is decomposed into an inventory of neural reasoning modules, which can be composed on the fly based on the user profile to execute $f_\mathrm{path}'$.
\end{itemize}

\section{Methodology}
Under the coarse-to-fine paradigm, we present a corresponding method called \approach{} to approach the problem of KG reasoning for recommendation.
As illustrated in Fig.~\ref{fig:method}, given a KG (a), a user profile is first composed to capture prominent user-centric patterns in the coarse stage (b). 
To conduct multi-hop path reasoning guided by the user profile, we decompose the reasoner $\phi$ into an inventory of neural reasoning modules (c).
In the fine stage, the selective neural symbolic reasoning modules are composed based on the user profile (d), which are exploited by a Profile-guided Path Reasoning (PPR) algorithm to efficiently perform batch path reasoning for recommendation (e).

\subsection{Coarse-Stage: User Profile Composition}\label{sec:coarse}
Given a user $u$, the goal of $f_\mathrm{profile}$ is to find a set of user-centric patterns that can distinctively characterize user behaviors, so that the potential paths with these patterns are more likely to arrive at items of interest to the given user.
Since e-commerce KGs usually contain a large number of relations, we first adopt an off-the-shelf random walk based algorithm \cite{lao2011random} to produce a candidate set of $M$ user-centric patterns, $\Pi=\{\pi_1,\pi_2,\ldots,\pi_M\}$, with maximum length $H$, from interacted user--item pairs in $\mathcal{G}$.
To compose the user profile, one naive way is to assign the weights in proportion to the frequency of these retrieved patterns.
However, this only provides overall information of user behavior towards items and is empirically shown not to achieve satisfying performance compared to personalized user profile (details in Section \ref{sec:exp_profile}).

\subsubsection{Personalized Pattern Selection}
The task of user profile composition now turns to selecting a subset from $\Pi$ and assigning weights that reflect the prominent behaviors for each user.
Formally, let $V_\Theta(u,\pi)$ be the prominence of a user-centric pattern $\pi$ for user $u$.
Intuitively, if $\pi$ is prominent with a larger value of $V_\Theta(u,\pi)$, it is more likely that the reasoning model $\phi$ can derive a path with pattern $\pi$ from $u$ to potential items of interest.
Hence, we define $V_\Theta(u,\pi)$ to be the likelihood of ``correct'' paths being generated by $\phi$:
\begin{equation}\label{eq:prominence}
V_\Theta(u,\pi) = \mathbb{E}_{L\sim D_\pi}[\log P_\Theta(L\mid u)],
\end{equation}
where $D_\pi$ denotes the set of paths with pattern $\pi$ between the user $u$ and interacted items in $\mathcal{G}_\mathrm{u}$, and $\log P_\Theta(L|u)$ is defined in Eq.~\ref{eq:prob}.
Here, we assume the reasoner $\phi$ has been trained and the parameter $\Theta$ is fixed.
The representation and model learning details will be discussed in Section \ref{sec:module}.

With the help of $V_\Theta(u,\pi)$, we propose a heuristic method to select prominent patterns to compose the profile for each user.
Specifically, the goal is to determine the weights $\{w_1,\ldots,w_M\}$ of candidate patterns in $\Pi$ and only the patterns with positive weights are kept.
This can be formalized as an optimization problem:
\begin{equation}\label{eq:compose}
\begin{aligned}
\max_{w_1,\ldots,w_M}\quad & \sum_j w_{j}\, V_\Theta(u,\pi_j) \\
\text{s.t.}\quad & \sum_j w_{j}=K,~ 0\le w_{j}\le K_j, j\in[M],
\end{aligned}
\end{equation}
where $K_j$ is the upper bound of the quantity of pattern $\pi_j$ to be adopted. 
The optimization problem corresponds to the well-studied bounded knapsack problem with equal weights $1$ and can be easily solved \cite{cormen2009introduction}.
Consequently, the user profile can be derived from Eq.~\ref{eq:compose} by $\mathcal{T}_u=\{(\pi_j,w_j)\mid \pi_j\in\Pi, w_j>0, j\in[M]\}$ (see example in Fig.~\ref{fig:method}(b)). 
Each positive $w_j$ specifies the number of paths with pattern $\pi_j$ to be generated by $f_\mathrm{path}$ (Section \ref{sec:fine}).

\subsubsection{Modularized Reasoning Model}\label{sec:module}
As introduced in Section \ref{sec:paradigm}, the reasoner $\phi$ parametrized by $\Theta$ determines the next-step decision in path-finding.
It maps the given user $u$ and historic trajectory $h_t$ to the conditional probability of choosing outgoing edge $(r_t,e_t)$, i.e., $\phi:u,h_t\mapsto P_\Theta(r_t,e_t|u,h_t)$.
Inspired by previous work \cite{xian2019kgrl}, we can treat $\phi$ as a stochastic policy network \cite{sutton2018reinforcement}.
However, instead of solving a reinforcement learning problem that requires a careful handcrafted design of good reward functions, we train the model $\phi$ via behavior cloning \cite{sutton2018reinforcement} by reusing the sampled paths that are previously retrieved to produce candidate patterns $\Pi$.

Nevertheless, learning $\phi$ is still challenging due to the huge search space in the KG, where the out-degrees of nodes can be very large and the number of connecting edges varies from node to node.
To address this, instead of representing $\phi$ as a deep and complex neural network to increase the reasoning capability, we propose to maintain an inventory of shallow \emph{neural symbolic reasoning modules} $\phi_r$ with parameter $\Theta_r$ for each relation $r$ in $\Pi$, as shown in Fig.~\ref{fig:method}(c).
Each $\phi_r(\mathbf{u},\mathbf{h};\Theta_r): \mathbb{R}^d\times\mathbb{R}^d\mapsto\mathbb{R}^d$ takes as input a user embedding $\mathbf{u}$ and a history embedding $\mathbf{h}$ and outputs the estimated vector of the next-hop entity.
The network structure of each $\phi_r$ is defined as:
\begin{equation}\label{eq:module}
\phi_r(u,h;\Theta_r) = \sigma(\sigma([\mathbf{u};\mathbf{h}]W_{r,1})W_{r,2})W_{r,3},
\end{equation}
where $[;]$ denotes concatenation, $\sigma(\cdot)$ is a nonlinear activation function (e.g., ReLU~\cite{relu}), and $\Theta_r=\{W_{r,1},W_{r,2},W_{r,3}\}$ are the learnable parameters for the module $\phi_r$.

With the module $\phi_{r_t}$, 
we can compute the probability
\begin{equation}\label{eq:onehop}
P_\Theta(r_t,e_t\mid u, h_t) \approx \frac{1}{Z} \exp(\langle \phi_{r_t}(\mathbf{u},\mathbf{h}_t;\Theta_{r_t}),\mathbf{e}_t \rangle),
\end{equation}
\noindent where $Z=\sum_{e_t'}\exp(\langle \phi_{r_t}(\mathbf{u},\mathbf{h}_t;\Theta_{r_t}),\mathbf{e}_t' \rangle)$ is the normalization term over all possible next-hop entities, and $\langle\cdot,\cdot \rangle$ is the dot product.

The benefits of this design are threefold.
First, the total number of parameters of maintaining such small modules is smaller than that of a deep and complex neural network.
Second, the space of next-hop actions is reduced from $(r_t,e_t)$ (all outgoing edges) to $e_t$ (only the edges of given relation), since the relation can be determined by the user profile.
Third, outputting a continuous vector can effectively solve the issue of varying numbers of outgoing edges.

\paragraph{Objectives}
We consider the set of all parameters $\Theta=\{\mathbf{e}|\forall e\in \mathcal{E}\}\cup\{\Theta_r|\forall r\in\Pi\}$, where $\mathbf{e}$ denotes the entity embedding and is initialized with a pretrained KG embedding \cite{bordes2013translating}.
Given a positive path $L=\{u,r_1,e_1,\ldots,e_{T-1},r_T,i^+\}$ with $(u,r_{ui},i^+)\in\mathcal{G}$, the behavior cloning aims to minimize the following loss over $\Theta$:
\begin{equation}\label{eq:loss_path}
\ell_\textrm{path}(\Theta;L) = -\log P_\Theta(L|u) = -\sum_{t=1}^{T} \log P_\Theta(r_t,e_t|u,h_t).
\end{equation}

However, the objective in Eq.~\ref{eq:loss_path} only forces the reasoning modules to fit the given path, but cannot identify which path may finally lead to potential items of interest.
Therefore, we impose an additional pairwise ranking loss $\ell_\mathrm{rank}(\Theta;L)$ to jointly train the parameters $\Theta$:
\begin{align}\label{eq:loss_rank}
\ell_\mathrm{rank}(\Theta;L) = -\mathbb{E}_{i^-\sim D_u^-}\left[\sigma\left(
\left\langle \mathbf{i}^+,\hat{\mathbf{e}}_{T} \right\rangle -
\left\langle \mathbf{i}^-,\hat{\mathbf{e}}_{T} \right\rangle
\right)\right],
\end{align}
where $D_u^-$ denotes the set of negative items of user $u$, i.e., $D_u^-=\{i|i\in\mathcal{I}, (u,r_{ui},i)\not\in \mathcal{G}\}$, $\hat{\mathbf{e}}_{T} = \phi_{r_T}(\mathbf{u},\mathbf{h}_T;\Theta_{r_T})$, and $\sigma(\cdot)$ is the sigmoid function.

By aggregating Eqs.~\ref{eq:loss_path} and \ref{eq:loss_rank} over all users in KG $\mathcal{G}_\mathrm{u}$, the overall goal is to minimize the following objective:
\begin{equation}\label{eq:loss_all}
\ell_\mathrm{all}\left(\Theta\right) = \sum_u \sum_{L\sim \mathcal{L}_u} \ell_\mathrm{path}(\Theta;L)+\lambda \ell_\mathrm{rank}(\Theta;L),
\end{equation}
where $\mathcal{L}_u=\{L_{u\leadsto i+}\mid (u,r_{ui},i^+)\in\mathcal{G}_\mathrm{u}, \text{pattern}(L_{u\leadsto i+})\in\Pi\}$, and $\lambda$ is the weighting factor to balance between the two losses.

\begin{algorithm}[t]
\caption{Profile-guided Path Reasoning (PPR) Algorithm}
\label{alg:program-exe}
\begin{algorithmic}[1]
\State \textbf{Input:} user $u$, user profile $\mathcal{T}_u$.
\State \textbf{Output:} $K$ user-centric paths.
\Procedure{Main}{$ $}
\State Construct layout tree $T_u$ based on user profile $\mathcal{T}_u$.
\State $x\gets \texttt{\scshape Root}(T_u)$,\quad $\hat{\mathbf{x}} \gets \mathbf{u}$,\quad $\mathcal{L}_x \gets \{\{u\}\}$.
\State Initialize queue $Q \gets \texttt{\scshape Children}(x)$.
\While{$Q \neq \emptyset$}
  \State $x \gets Q.pop()$,\quad $p \gets \texttt{\scshape Parent}(x)$.
  \State $\hat{\mathbf{x}} \gets \phi_{r_{x}}(\mathbf{u},\hat{\mathbf{p}};\Theta_{r_{x}})$.
  \State Initialize $\mathcal{L}_x \gets \{\}$.
  \For{$L \in \mathcal{L}_p$}
    \State $E_x \gets \{e' \mid \forall (e_{|L|},r_{x},e')\in \mathcal{G}$, $\tau(e')=\tau_t(r_{x}), \mathrm{rank}(\langle \hat{\mathbf{x}}, \mathbf{e}' \rangle) \le n_x \}$.
    \State $\mathcal{L}_x \gets \mathcal{L}_x \cup (L \cup \{e'\})$, for $e'\in E_x$.
  \EndFor
  \State Update $Q \gets Q\cup \texttt{\scshape Children}(x)$.
\EndWhile
\State \Return $\bigcup_{x\in \texttt{\scshape Leaves}(T_u)} \mathcal{L}_x$.
\EndProcedure
\end{algorithmic}
\end{algorithm}

\subsection{Fine-Stage: Path Reasoning for Recommendation}\label{sec:fine}
Given the composed user profile $\mathcal{T}_u=\{(\pi_1,w_1),\ldots,(\pi_M,w_M)\}$ of user $u$, the goal of $f_\mathrm{path}$ is to output $K$ reasoning paths along with items such that the number of paths with pattern $\pi_j$ is proportional to $w_j$.
Considering that finding each path individually is inefficient due to repeated node visitation and calculation \cite{xian2019kgrl}, we propose a \emph{Profile-guided Path-Reasoning} algorithm (PPR) that is capable of finding a batch of paths simultaneously via selective neural symbolic reasoning modules according to the composed user profile.
As illustrated in Fig.~\ref{fig:method}(d), it first constructs a layout tree $T_u$ from the user profile $\mathcal{T}_u$ to specify the execution order of neural symbolic reasoning modules.
Then, the reasoning modules are executed level by level to produce the next-hop embeddings that are employed to find the closest entities in the KG (Fig.~\ref{fig:method}(e)).

The details of the algorithm are given in Alg.~\ref{alg:program-exe}.
Specifically, the layout tree $T_u$ (line 4) is first constructed by merging patterns in $\mathcal{T}_u$, so that each node $x\in T_u$ is associated with a relation $r_x$ (a dummy relation is used for the root node), and each root-to-leaf tree path corresponds to a pattern in $\mathcal{T}_u$.
Next, an integer $n_x$ is assigned to each node $x$, which specifies the number of entities to be generated at the current position.
If node $x$ is the root, one sets $n_x=1$. If $x$ is a leaf, $n_x$ is initialized with $w_j$, i.e., the weight of pattern $\pi_j$ that ends with relation $r_x$. Otherwise, $n_x$ is updated by $n_x=\min_{c\in\mathrm{children}(x)}(n_c)$, and subsequently, the value at each child $c$ of node $x$ will be refreshed as $n_c'=\lfloor n_c/n_x \rfloor$.

In fact, $T_u$ specifies the layout of a tree-structured neural network composed of reasoning modules $\phi_{r_x}$ at each node $x$ with relation $r_x$. 
The execution process of the network is described in Alg.~\ref{alg:program-exe} (lines 5-15) to derive $K$ reasoning paths simultaneously.
It starts at the root node of $T_u$ and follows level-order traversal to generate paths.
At each node $x\in T_u$, $\phi_{r_x}$ takes as input the user embedding $\mathbf{u}$ and the embedding from its parent node and outputs an embedding vector denoted by $\hat{\mathbf{x}}$.
Meanwhile, a set of new paths $\mathcal{L}_x$ up to node $x$ is generated based on $\hat{\mathbf{x}}$ as well as the paths from its parent node $\mathcal{L}_p$.
Specifically, for each path $L\in\mathcal{L}_p$, we find at most $n_x$ new entities such that each of them is connected to the last entity in $L$ in the KG, and its embedding is most similar to $\hat{\mathbf{x}}$.
Eventually, we obtain the final results by aggregating all the paths at the leaf nodes and rank them based on the dot-product score in Eq.~\ref{eq:onehop}.

\begin{table}[t]
\begin{adjustbox}{width=1.0\linewidth}
\begin{tabular}{@{}lllll@{}}
\toprule
& \textbf{CDs \& Vinyl} & \textbf{Clothing} & \textbf{Cell Phones} & \textbf{Beauty} \\
\midrule
\#Users    & 75,258  & 39,387  & 27,879  & 22,363 \\
\#Items    & 64,443  & 23,033  & 10,429  & 12,101 \\
\#Interactions & 1.10M   & 278.86K & 194.32K & 198.58K \\
\#Entities & 581,105 & 425,534 & 163,255 & 224,080 \\
\#Relations & 16 & 16 & 16 & 16 \\
\#Triples & 387.43M & 36.37M & 37.01M & 37.73M \\
\bottomrule
\end{tabular}
\end{adjustbox}
\caption{Statistics of four real-world Amazon KG datasets: \emph{CDs \& Vinyl}, \emph{Clothing}, \emph{Cell Phones}, and \emph{Beauty}.
\vspace{-20pt}
}
\label{tab:stats}
\end{table}

\begin{table*}[t]
\begin{adjustbox}{width=\textwidth}
\begin{tabular}{@{}llcccclcccclcccclcccc@{}}
\toprule
&& \multicolumn{4}{c}{\textbf{CDs \& Vinyl}} && \multicolumn{4}{c}{\textbf{Clothing}} && \multicolumn{4}{c}{\textbf{Cell Phones}} && \multicolumn{4}{c}{\textbf{Beauty}} \\
\cmidrule{3-6} \cmidrule{8-11} \cmidrule{13-16} \cmidrule{18-21}
Measures ($\%$) && NDCG  & Recall & HR    & Prec. && NDCG  & Recall & HR   & Prec. && NDCG  & Recall & HR    & Prec. && NDCG  & Recall & HR     & Prec. \\
\midrule 
BPR            && 2.009 & 2.679 & 8.554  & 1.085 && 0.601 & 1.046 & 1.767 & 0.185 && 1.998 & 3.258 & 5.273  & 0.595 && 2.753 & 4.241  & 8.241  & 1.143\\
BPR-HFT        && 2.661 & 3.570 & 9.926  & 1.268 && 1.067 & 1.819 & 2.872 & 0.297 && 3.151 & 5.307 & 8.125  & 0.860 && 2.934 & 4.459  & 8.268  & 1.132\\
DeepCoNN       && 4.218 & 6.001 & 13.857 & 1.681 && 1.310 & 2.332 & 3.286 & 0.229 && 3.636 & 6.353 & 9.913  & 0.999 && 3.359 & 5.429  & 9.807  & 1.200\\
CKE            && 4.620 & 6.483 & 14.541 & 1.779 && 1.502 & 2.509 & 4.275 & 0.388 && 3.995 & 7.005 & 10.809 & 1.070 && 3.717 & 5.938  & 11.043 & 1.371 \\
RippleNet      && 4.871 & 7.145 & 15.727 & 1.852 && 2.195 & 3.892 & 6.032 & 0.603 && 4.837 & 7.716 & 11.454 & 1.101 && 5.162 & 8.127 & 14.681 & 1.699 \\
KGAT           && 5.411 & 7.764 & 17.173 & 2.120 && 3.021 & 5.172 & 7.394 & 0.747 && 5.111 & 8.978 & 12.589 & 1.296 && 6.108 & 10.022 & 16.740 & 1.893 \\
HeteroEmbed    && 5.563 & \underline{7.949} & \underline{17.556} & \underline{2.192} && \underline{3.091} & \underline{5.466} & \underline{7.972} & \underline{0.763} && \underline{5.370} & \underline{9.498} & \underline{13.455} & \underline{1.325} && \underline{6.399} & \underline{10.411} & \underline{17.498} & \underline{1.986} \\
PGPR           && \underline{5.590} & 7.569 & 16.886 & 2.157 && 2.858 & 4.834 & 7.020 & 0.728 && 5.042 & 8.416 & 11.904 & 1.274 && 5.449 & 8.324 & 14.401 & 1.707 \\
\approach\ (Ours)  && \textbf{6.868} & \textbf{9.376} & \textbf{19.692} & \textbf{2.562} && \textbf{3.689} & \textbf{6.340} & \textbf{9.275} & \textbf{0.975} && \textbf{6.313} & \textbf{11.086} & \textbf{15.531} & \textbf{1.692} && \textbf{7.061} & \textbf{10.948} & \textbf{18.099} & \textbf{2.270} \\
\midrule
Improvement (\%)    && +22.86 & +17.95 & +12.17 & +16.88 && +19.34 & +15.99 & +16.34 & +24.52 && +17.56 & +16.72 & +15.43 & +24.60 && +10.34 & +5.16 & +3.43 & +14.07 \\
\bottomrule
\end{tabular}
\end{adjustbox}
\caption{Overall recommendation performance of our method compared to other approaches on four benchmarks. The results are computed based on top-10 recommendations in the test set and are given as percentages (\%). The best results are highlighted in bold font and the best baseline results are underlined.}
\vspace{-20pt}
\label{tab:eval}
\end{table*}

\subsection{Model Analysis}
For each user, the time complexity of PPR in Alg.~\ref{alg:program-exe} is $O(MH(Q + KdD))$, where $Q$ is the running time for executing each neural symbolic reasoning module, $d$ is the dimensionality of entity embeddings, $D$ is the maximum node degree in the KG.
Intuitively, there are at most $O(MH)$ nodes in $T_u$, and for each node, it costs $O(MHQ)$ time for the inference (forward pass) of the neural reasoning module, and $O(KdD)$ time to find nearest entities in Alg.~\ref{alg:program-exe}. 
Unlike existing methods \cite{xian2019kgrl,ai2018learning} that find each individual path separately, our PPR algorithm can derive all $K$ paths simultaneously in the tree level order.
If some resulting paths share the same entities, their corresponding embeddings will be computed only once and hence redundant computations are avoided.
The efficiency of the algorithm is also empirically evaluated in Section \ref{sec:time}.

\section{Experiments}\label{sec:experiment}
In this section, we extensively evaluate our proposed approach, providing a series of quantitative as well as qualitative analyses on several real-world datasets.

\subsection{Experimental Setup}
\subsubsection{Dataset}
We experiment on four domain-specific e-commerce datasets from Amazon \cite{he2016ups}, namely \emph{CDs and Vinyl}, \emph{Clothing}, \emph{Cell Phones}, and \emph{Beauty}. 
They provide both rich meta-information of products and diverse user behavior records such as purchase history, ratings, product reviews, and preferred styles.
Each dataset is considered as an individual benchmark that constitutes a user-centric KG with various types of relations (including inverse relations), which implies that results are not necessarily comparable across different domains.
Table \ref{tab:stats} summarizes the statistical information of the four datasets.
We adopt the same training (70\%) and test splits (30\%) as previous work \cite{ai2018learning,xian2019kgrl},
which are publicly available\footnote{\url{https://github.com/orcax/PGPR}}.

\subsubsection{Baselines and Metrics}
We consider three categories of recommendation approaches as baselines in the following experiments.
\begin{itemize}[leftmargin=*]
\item MF-based models:
    \textbf{BPR} \cite{rendle2009bpr} is a Bayesian personalized method that optimizes a pairwise ranking between different user--item pairs for top-$N$ recommendation.
    \textbf{BPR-HFT} \cite{mcauley2013hidden} is a review-based recommendation method based on Hidden Factors and Topics (HFT) to learn latent representations of users and items with the topic distributions incorporated.
    \textbf{DeepCoNN} \cite{zheng2017joint} makes recommendations through a Deep Cooperative Neural Network based on reviews, which is capable of encoding both users and products for rating prediction.
\item KG embedding models:
    \textbf{CKE} \cite{zhang2016collaborative}, or Collaborative Knowledge base Embedding, is a neural recommendation method based on jointly integrating matrix factorization and heterogeneous graph data to infer recommendations.
    \textbf{RippleNet}~\cite{wang2018ripplenet} incorporates a KG into recommendation by propagating user preferences on entities.
    \textbf{KGAT} \cite{wang2019kgat} is the the state-of-the-art  KG-based model using graph-based attention techniques.
\item Path reasoning models: 
    \textbf{HeteroEmbed} \cite{ai2018learning} is the state-of-the-art Rec-First approach based on TransE~\cite{bordes2013translating} embeddings for recommendations, followed by a post-hoc graph search to find paths.
    \textbf{PGPR} \cite{xian2019kgrl} is the state-of-the-art path-guided model, which conducts path reasoning using reinforcement learning. %
\end{itemize}
For all models, we adopted the same metrics as previous work \cite{xian2019kgrl} to evaluate the top-10 recommendations of each user in the test set, including
Normalized Discounted Cumulative Gain (\textbf{NDCG}), \textbf{Recall}, Hit Rate (\textbf{HR}), and Precision (\textbf{Prec.}).

\subsubsection{Implementation Details}

In our model, the entity embedding dimensionality is $100$. 
In each neural relation module $\phi_r$ with respect to some relation $r$, the parameters are
$W_{r,1}\in\mathbb{R}^{200\times 256}$,
$W_{r,2}\in\mathbb{R}^{256\times 256}$, and
$W_{r,3}\in\mathbb{R}^{256\times 100}$.
We use Xavier initialization for the parameters and train them with  Adam optimization \cite{kingma2014adam} with a learning rate of $10^{-4}$, batch size of 128, and a number of training epochs of 20.
The history $h_t$ is set to $e_{t-1}$.
The weighting factor $\lambda$ for the ranking loss is set to 10. The number of output paths $K$ is $15$.
For fair comparison with previous work \cite{ai2018learning,xian2019kgrl}, we also restrict the maximum path length $H$ to 3, which leads to 15 candidate user-centric patterns in $\Pi$.
The influence of these hyperparameters will be studied in Section \ref{sec:ablation}.

\subsection{Overall Performance}
We first show the top-10 recommendation performance of our proposed method \approach{} compared to all baselines.
We evaluate each setting 5 times and report the average scores in Table \ref{tab:eval}.

Overall, we observe that our method outperforms three kinds of state-of-the-art methods (KGAT, HeteroEmbed, PGPR) by a large margin across all settings.
For example, on the Clothing dataset, our model achieves 6.340\% in Recall, which is higher than 5.172\% by KGAT, 5.466\% by HeteroEmbed, and 4.834\% of PGPR. 
Similar trends can also be observed on other benchmarks.
Additionally, our model shows better ranking performance than the baselines in terms of NDCG.
This is mainly attributed to the ranking loss in Eq.~\ref{eq:loss_rank}, which encourages the model to identify the path based on whether it can lead to good items. 
The influence of the ranking loss will be studied in Section \ref{sec:rank}.

Note that KG embedding based approaches such as RippleNet and KGAT are less competitive on these datasets.
One possible reason is that unlike KGs such as Freebase, where the reasoning rules are objective and explicit (e.g., \emph{HasNationality = BornIn $\wedge$ CityIn}), the patterns of user behavior towards items are more diverse and uncertain in e-commence settings
(e.g., many factors can contribute to a user purchase behavior), making it harder to mine useful information.
Our coarse-to-fine method can first learn a sketch of user behavior (i.e., user profile), which filters out noisy information that may be irrelevant to conduct path reasoning. That is why our model is able to achieve better recommendation performance.
The effectiveness of user profiles is studied in the next section.

\subsection{Effectiveness of User Profile (Coarse-Stage)}\label{sec:exp_profile}
\begin{table}[h]
\centering
\begin{adjustbox}{width=\linewidth}
\begin{tabular}{@{}lcccccccccc@{}}
\toprule
&& \multicolumn{4}{c}{\textbf{CDs \& Vinyl}}  && \multicolumn{4}{c}{\textbf{Clothing}} \\
\cmidrule{3-6} \cmidrule{8-11}
    && NDCG  & Recall & HR    & Prec.  && NDCG  & Recall & HR    & Prec. \\
\midrule
PGPR      && 5.590 & 7.569 & 16.886 & 2.157 && 2.858 & 4.834 & 7.020 & 0.728 \\
Rand      && 5.308 & 7.217 & 16.158 & 2.003 && 2.654 & 4.727 & 6.875 & 0.680 \\
Prior     && 5.924 & 8.259 & 17.825 & 2.327 && 3.157 & 5.031 & 7.376 & 0.773  \\
Ours      && \textbf{6.868} & \textbf{9.376} & \textbf{19.692} & \textbf{2.562} && \textbf{3.689} & \textbf{6.340} & \textbf{9.275} & \textbf{0.975} \\
\midrule
&& \multicolumn{4}{c}{\textbf{Cell Phones}}  && \multicolumn{4}{c}{\textbf{Beauty}} \\
\cmidrule{3-6} \cmidrule{8-11}
    && NDCG  & Recall & HR    & Prec.  && NDCG  & Recall & HR    & Prec. \\
\midrule
PGPR      && 5.042 & 8.416 & 11.904 & 1.274 && 5.449 & 8.324  & 14.401 & 1.707 \\
Rand      && 4.545 & 7.229 & 10.192 & 1.087 && 5.293 & 8.256  & 14.564 & 1.718 \\
Prior     && 5.255 & 9.842 & 13.097 & 1.359 && 6.180 & 9.393 & 16.258 & 2.024 \\
Ours      && \textbf{6.313} & \textbf{11.086} & \textbf{15.531} & \textbf{1.692} && \textbf{7.061} & \textbf{10.948} & \textbf{18.099} & \textbf{2.270} \\
\bottomrule
\end{tabular}
\end{adjustbox}
\caption{Results of recommendation performance using different user profile variants.}
\label{tab:program-exp}
\vspace{-20pt}
\end{table}

In this experiment, we evaluate the effectiveness of the approach to compose user profiles as described in Section \ref{sec:coarse}.
Specifically, we consider the following ways to compose different $\mathcal{T}_u$ for user $u$ while keeping the same path reasoning algorithm in Section \ref{sec:fine}.
\begin{itemize}[leftmargin=*]
	\item \emph{Rand} stands for randomly sampling a subset of patterns from $\Pi$ to compose $\mathcal{T}_u$. This straightforward method can represent the path reasoning methods without considering user profiles.
	\item \emph{Prior} samples the patterns from $\Pi$ proportional to their frequencies and discards low frequency patterns. This is equivalent to assigning each user the same profile based on global information.
	\item \emph{\approach{}} is the approach we propose, which estimates the weights by solving the optimization problem in Eq.~\ref{eq:compose}.
\end{itemize}
Additionally, we also compare to the SOTA path reasoning approach \emph{PGPR} that also fails to model user profiles.

The results on all datasets are reported in Table \ref{tab:program-exp}.
We observe that our model \approach{} with composed user profile exhibits better recommendation performance than other baselines.
This shows that the path reasoning guided by the user profile can find user-centric paths of higher quality, which are more likely to arrive at an item node of interest to the user.
In addition, we also note that the profile-driven methods \approach{} and \emph{Prior} outperform the ones without profiles (\emph{PGPR}, \emph{Rand}).
This suggests that user profiles can benefit the path reasoning process.

\subsection{Efficiency of Path Reasoning (Fine-Stage)}\label{sec:time}
\begin{table}[h]
\centering
\begin{adjustbox}{width=\linewidth}
\begin{tabular}{@{}lcrrcrr@{}}
\toprule
&& \multicolumn{2}{c}{\textbf{CDs \& Vinyl}}  && \multicolumn{2}{c}{\textbf{Clothing}} \\
\cmidrule{3-4} \cmidrule{6-7}
Time (s)  && Rec. (1k users) & Find (10k paths)  && Rec. (1k users) & Find (10k paths) \\
\midrule
PGPR       && $287.158\pm 5.213$ & $26.725\pm 0.572$ && $236.118\pm 4.840$ & $21.889\pm 0.437$ \\
Hetero.    && $53.984\pm 1.201$  & $21.674\pm 0.498$ && $55.482\pm 1.703$  & $18.492\pm 0.399$ \\
Indiv. && $71.769\pm 1.366$  & $25.229\pm 0.482$ && $61.519\pm 1.966$  & $20.128\pm 0.377$ \\
Ours       && $\mathbf{27.184\pm 1.026}$ & $\mathbf{17.851\pm 0.364}$ && $\mathbf{22.850\pm 1.378}$ & $\mathbf{15.233\pm 0.309}$  \\
\midrule
&& \multicolumn{2}{c}{\textbf{Cell Phones}}  && \multicolumn{2}{c}{\textbf{Beauty}} \\
\cmidrule{3-4} \cmidrule{6-7}
Time (s)  && Rec. (1k users) & Find (10k paths)  && Rec. (1k users) & Find (10k paths) \\
\midrule
PGPR       && $279.780\pm 5.135$ & $25.382\pm 0.563$ && $292.447\pm 6.139$ & $26.396\pm 0.591$ \\
Hetero.    && $48.125\pm 1.148$  & $20.037\pm 0.496$ && $51.392\pm 1.369$  & $21.492\pm 0.467$  \\
Indiv. && $62.259\pm 1.171$  & $23.735\pm 0.502$ && $68.158\pm 1.209$  & $24.938\pm 0.473$ \\
Ours       && $\mathbf{23.387\pm 1.124}$  & $\mathbf{15.591\pm 0.406}$ && $\mathbf{25.220\pm 1.141}$  & $\mathbf{16.813\pm 0.458}$ \\
\bottomrule
\end{tabular}
\end{adjustbox}
\caption{Time costs of recommendations per 1k users and path finding per 10k paths.}
\label{tab:time-exp}
\vspace{-20pt}
\end{table}

We further study the efficiency of our path reasoning algorithm in Section \ref{sec:fine} compared to other path-finding baselines.
Specifically, we consider the SOTA Path-Guided method \emph{PGPR} and the SOTA Rec-First method \emph{HeteroEmbed}.
We also include a variant of our algorithm in Alg.~\ref{alg:program-exe} named \emph{Indiv.}, which simply finds each individual path one by one.
These algorithms are evaluated on the empirical running time of 1) making recommendations (including both items and paths) for 1k users and 2) the path-finding process (only paths) for generating 10k paths.
All experiments are conducted on the same hardware with Intel i7-6850K CPU, 32G memory and one Nvidia 1080Ti GPU.
The results are reported in Table \ref{tab:time-exp}.

We observe that our method costs the least time for both tasks among all tested algorithms.
In particular, our method is about $10\times$ faster than \emph{PGPR} in making recommendations on all benchmarks, both of which aim to find paths with unknown destination.
One reason is that \emph{PGPR} is required to find a lot of candidate paths, which are then ranked to obtain top 10 paths for recommendation.
On the contrary, our method seeks out useful paths based on the user profile, and hence it saves much more time in path-reasoning based recommendation.
In addition, for both tasks, our method costs less time than \emph{Indiv.}, which means that the batch path finding algorithm in Alg.~\ref{alg:program-exe} is more efficient than finding paths individually.
Our algorithm thus avoids redundant computation of embeddings and nearest nodes searches.

\subsection{Robustness to Unseen Patterns}\label{sec:unseen}
\begin{table}[h]
\centering
\resizebox{1.0\linewidth}{!}{
\begin{tabular}{@{}lcrrrrcrrrr@{}}
\toprule
&& \multicolumn{4}{c}{\textbf{CDs \& Vinyl}}  && \multicolumn{4}{c}{\textbf{Clothing}} \\
\cmidrule{3-6} \cmidrule{8-11}
\#Patterns && NDCG   & Recall & HR     & Prec.  && NDCG  & Recall & HR    & Prec. \\
\midrule
100\%     && 6.868  & 9.376 & 19.692 & 2.562  && 3.689 & 6.340 & 9.275 & 0.975 \\
70\%      && 6.713  & 9.152 & 19.270 & 2.488  && 3.586 & 6.175 & 9.020 & 0.946 \\
Decrease  && 2.31\% & 2.45\% & 2.19\% & 2.98\% && 2.87\% & 2.68\% & 2.83\% & 3.05\% \\
\midrule
&& \multicolumn{4}{c}{\textbf{Cell Phones}}  && \multicolumn{4}{c}{\textbf{Beauty}} \\
\cmidrule{3-6} \cmidrule{8-11}
\#Patterns && NDCG   & Recall & HR     & Prec.  && NDCG  & Recall & HR    & Prec. \\
\midrule
100\%     && 6.313  & 11.086 & 15.531 & 1.692  && 7.061 & 10.948 & 18.099 & 2.270 \\
70\%       && 6.143  & 10.764 & 15.098 & 1.639  && 6.974 & 10.803 & 17.835 & 2.225 \\
Decrease   && 2.77\% & 2.99\% & 2.87\% & 3.23\% && 1.25\% & 1.34\% & 1.48\% & 2.02\% \\
\bottomrule
\end{tabular}
}
\caption{Experimental results for unseen patterns}
\vspace{-10pt}
\label{tab:unseen-exp}
\end{table}

Recall that the candidate set $\Pi$ cannot exhaustively cover all possible user-centric patterns in a very large KG, since only high frequency patterns will be collected by the algorithm \cite{lao2011random}.
Therefore, in this experiment, we investigate if unseen patterns that do not exist in $\Pi$ will influence the performance.
To conduct the experiment, in the coarse stage of user profile composition, we randomly preserve 70\% of the candidate patterns in $\Pi$ for each user to compose the user profile. 
The remaining 30\% of patterns are unseen to the model for each user. 
All other settings remain the default ones.

The results on the four datasets are reported in Table \ref{tab:unseen-exp}. 
It is interesting to see that the decrease in performance is at around 1.5--3\%, which is marginal compared to the regular setting. This shows that our model is robust to unseen patterns for user profile composition and can still provide high-quality recommendations.

\subsection{Ablation Study}\label{sec:ablation}
We study how different settings of hyperparameters influence the recommendation quality of our model.
We consider the ranking weight and the number of sampling paths on the Cell Phones dataset only due to space constraints.

\subsubsection{Influence of ranking loss}\label{sec:rank}
\begin{figure}[t]
\centering
\small
\hspace{-8pt}
\includegraphics[width=1.15in,height=0.92in]{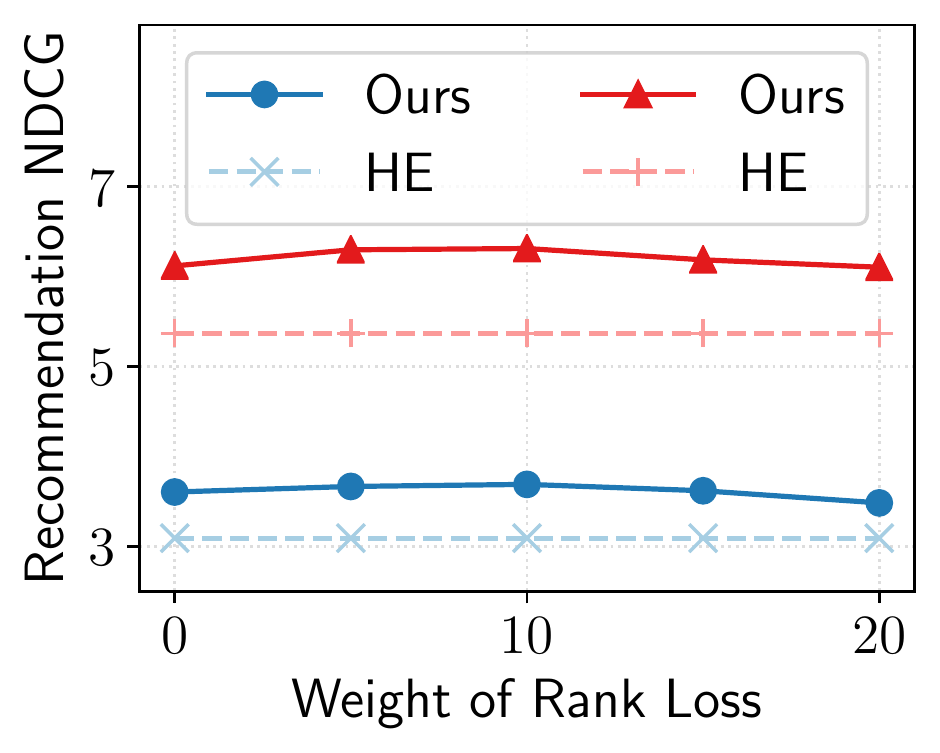}
\hspace{-5pt}
\includegraphics[width=1.15in,height=0.92in]{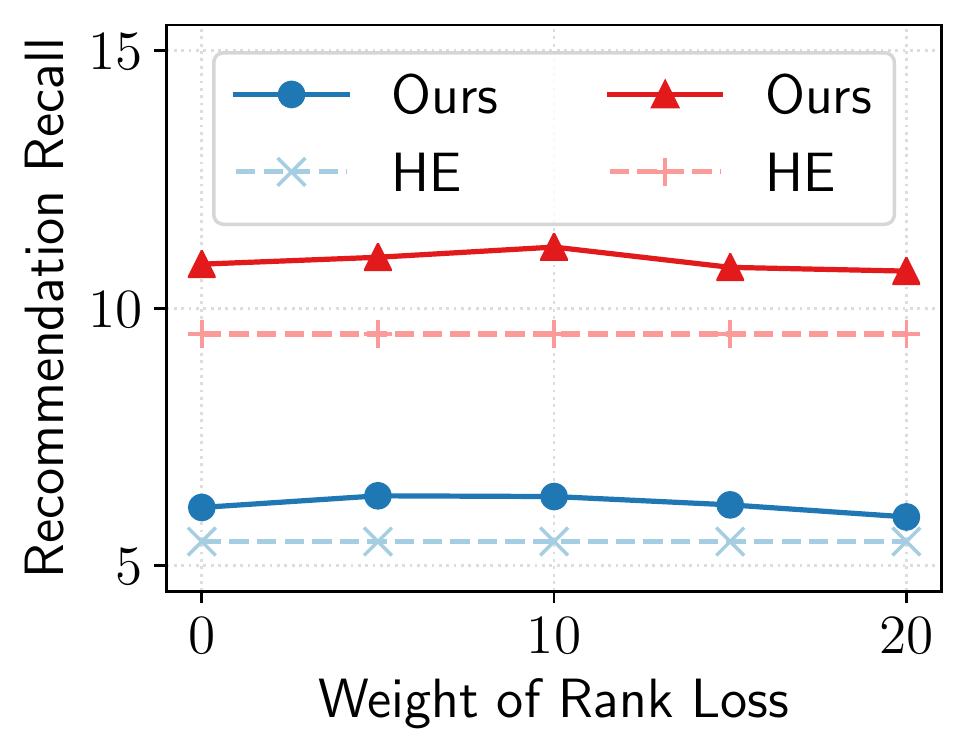}
\hspace{-5pt}
\includegraphics[width=1.15in,height=0.92in]{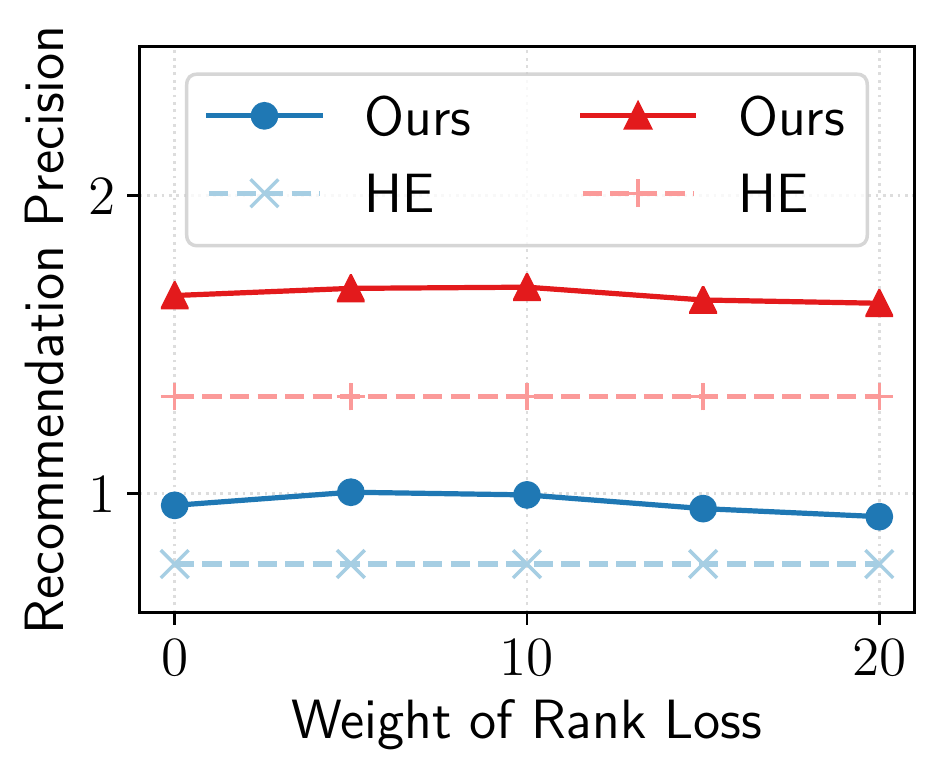} \\
\vspace{-3pt}
\hspace*{0.12in} {(a) NDCG \hspace{0.62in} (b) Recall \hspace{0.62in} (c) Precision }
\vspace{-8pt}
\caption{Results of varying ranking weights on Clothing (blue) and Cell Phones (red) datasets. (HE: \cite{ai2018learning})}
\label{fig:rank_weight}
\vspace{-10pt}
\end{figure}

We first show the influence of the ranking loss in Eq.~\ref{eq:loss_all} under different values of the weighting factor $\lambda \in \{0, 5, 10, 15, 20\}$, where $\lambda=0$ means no ranking loss is imposed for training.
The results are plotted in Fig.~\ref{fig:rank_weight}, including our model (red curves) and the best baseline HeteroEmbed (blue curves).

We observe two interesting trends.
First, our model consistently outperforms HeteroEmbed under all settings of $\lambda$ in terms of NDCG, Recall, and Precision. Even without the ranking loss, our model can still guarantee a high quality of recommendation. On the other hand, a proper choice of $\lambda$ (e.g., $\lambda=10$) not only benefits the direct ranking effect (NDCG), but also boosts the model's ability to find more relevant items (recall, hit rate, and precision).
Second, a larger weight of the ranking loss may not always entail a better performance, since there is a trade-off between the ranking (Eq.~\ref{eq:loss_rank}) and path regularization (Eq.~\ref{eq:loss_path}). 
This is reasonable because if the ranking loss plays a dominant role, which implies that the model pays less attention to fitting paths, as a consequence, it may fail to find the correct paths that reach promising items. %

\begin{figure}[t]
\centering
\small
\hspace{-8pt}
\includegraphics[width=0.87in]{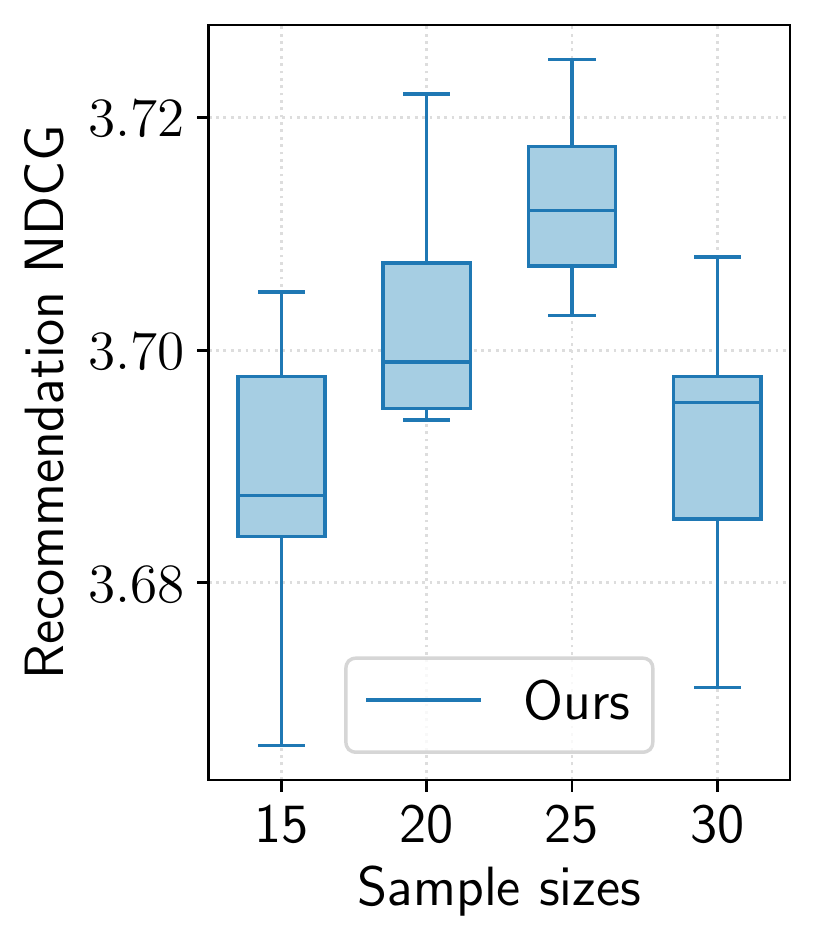}
\hspace{-5pt}
\includegraphics[width=0.87in]{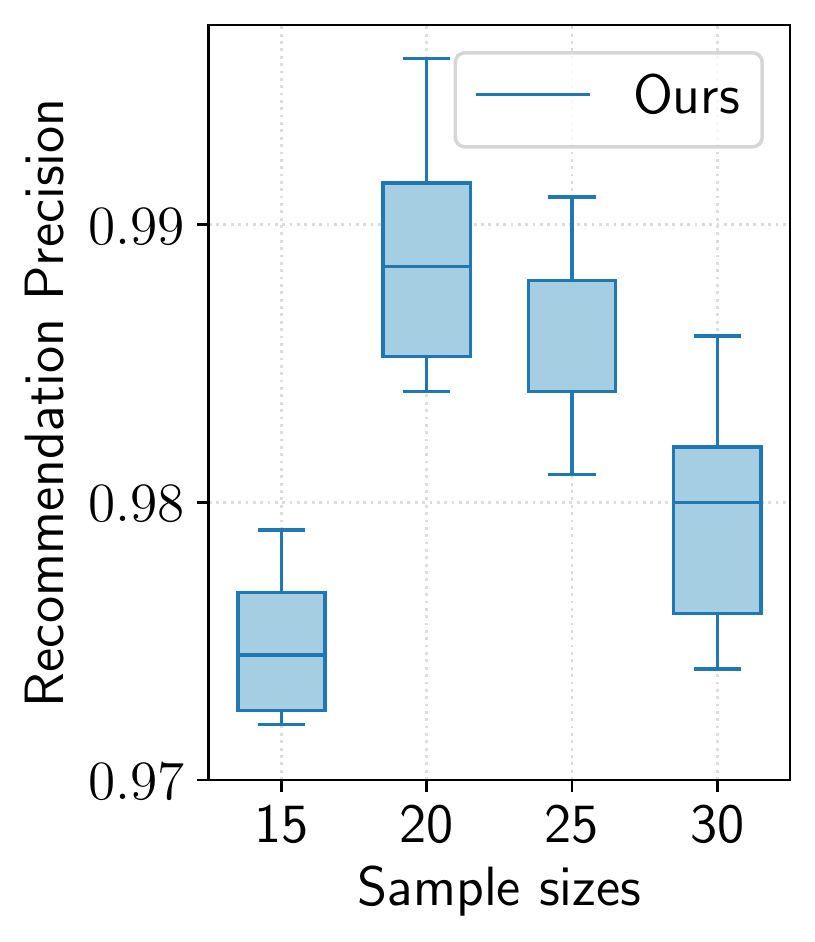}
\hspace{-5pt}
\includegraphics[width=0.87in]{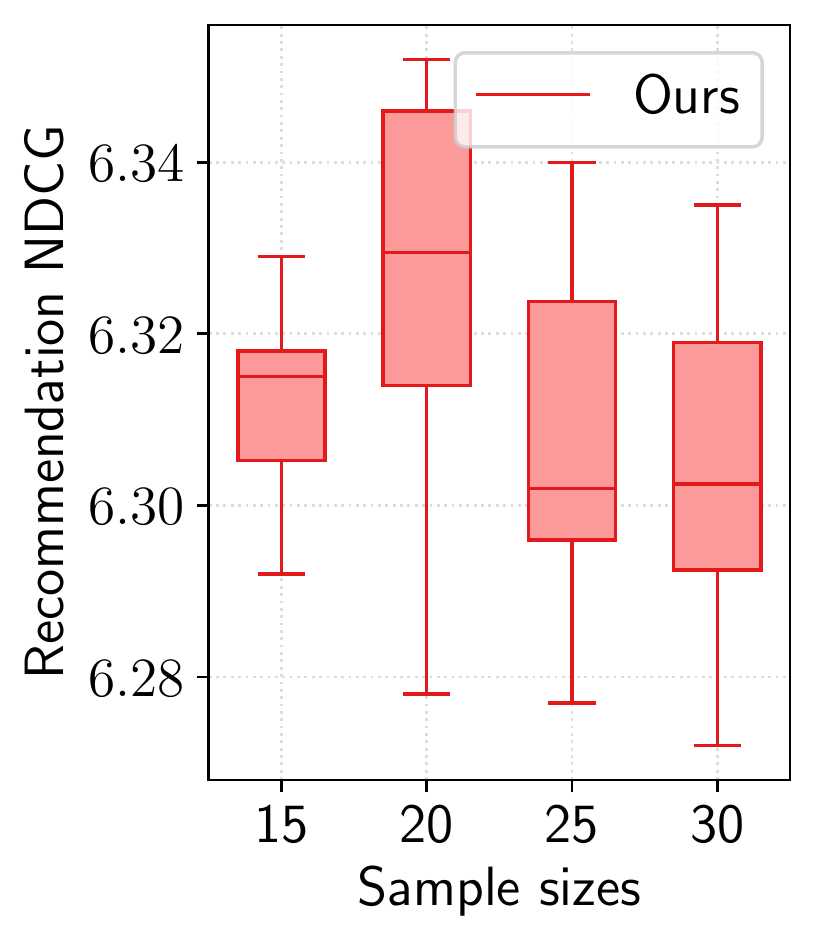}
\hspace{-5pt}
\includegraphics[width=0.87in]{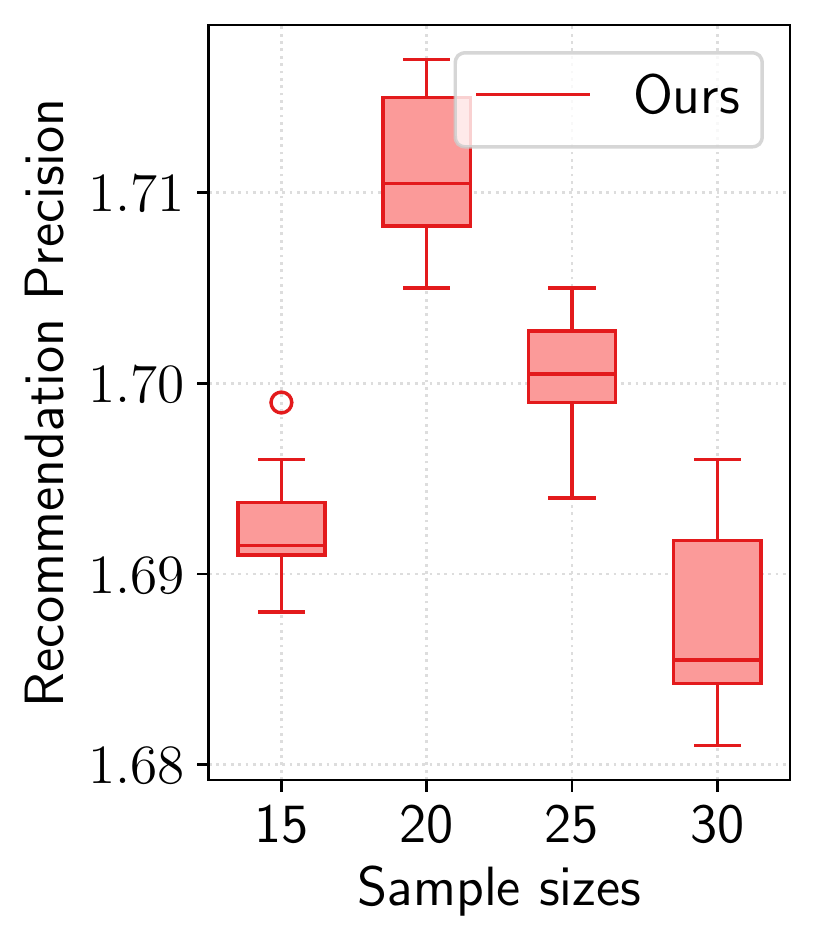} \\
\vspace{-3pt}
\hspace*{0.10in} {(a) NDCG \hspace{0.3in} (b) Precision \hspace{0.3in} (c) NDCG \hspace{0.3in} (d) Precision}
\vspace{-6pt}
\caption{Results of different number of output reasoning paths on Clothing (blue) and Cell Phones (red) datasets.}
\label{fig:output_paths}
\vspace{-10pt}
\end{figure}

\begin{figure*}[ht]
\centering
\includegraphics[width=0.92\textwidth]{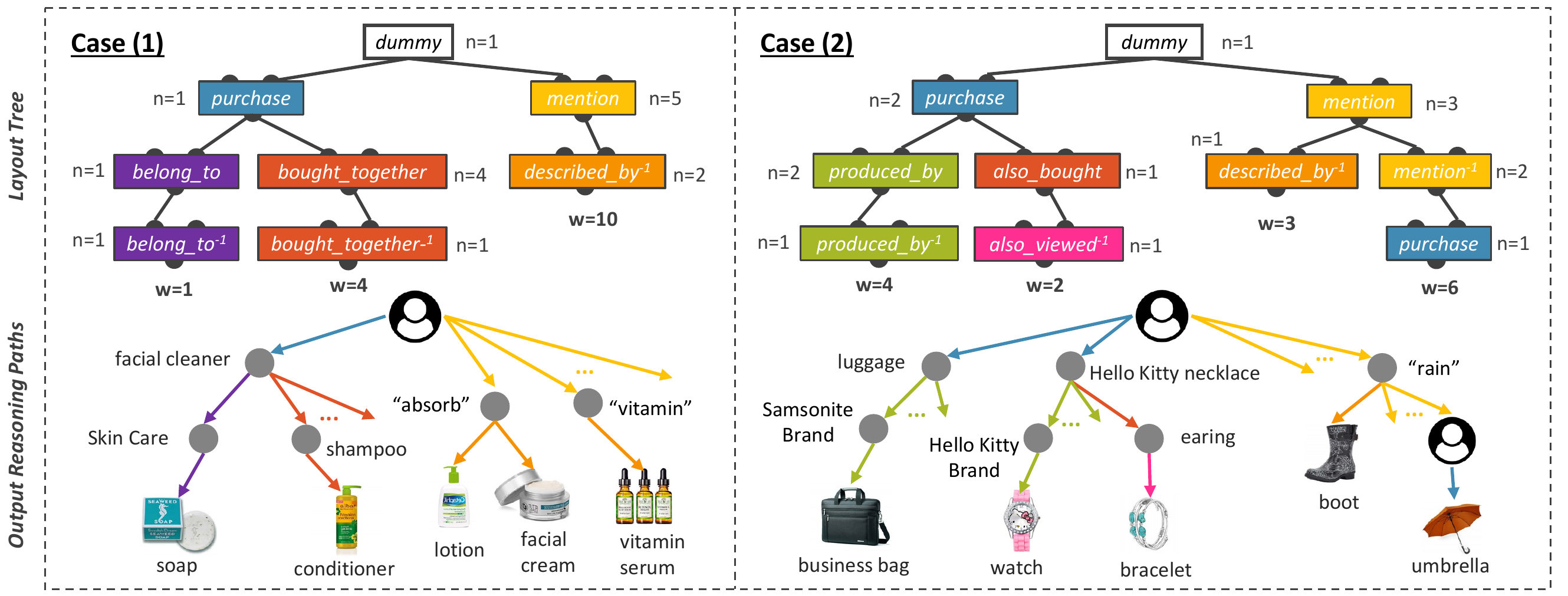}
\vspace{-10pt}
\caption{Two real cases discovered by our model, each containing a layout tree merged from user profile and a subset of reasoning paths. The end nodes in the resulting paths are the predicted items for recommendation.}
\label{fig:cases}
\end{figure*}

\subsubsection{Influence of sampling sizes of output paths}
Furthermore, we study how the performance varies with different sampling sizes of output reasoning paths $K\in \{15, 20, 25, 30\}$ (see Section \ref{sec:fine}).

In Fig.~\ref{fig:output_paths}, we illustrate with box plots the recommendation performance of all users in terms of various metrics.
We observe similar trends across all metrics in that there exists an optimal choice of $K$ under each setting, e.g., $K=20$ for NDCG on the Cell Phones dataset.
The variances are within acceptable ranges, which means that the path reasoning procedure of our model leads to satisfying results for most of the users.
One possible reason is that some items suitable for recommendation are in fact ranked relatively low. Smaller sampling sizes lead to smaller search spaces that preclude the discovery of such low-ranked items.

\subsection{Case Study}
We showcase two recommendation examples with path-based explanations produced by our model \approach{}.
As shown in Fig.~\ref{fig:cases}, each case consists of a layout tree merged from the user profile along with a subset of generated reasoning paths.
In Case 1, the pattern containing the ``mention'' relation takes the dominant role ($w=10$). For example, the user mentions the keywords ``absorb'', ``vitamin''. The recommended items ``lotion'' and ``facial cream'' match ``absorb'', and ``vitamin serum'' is also consistent with ``vitamin''. %
Case 2 shows a user profile with more diverse patterns. For example, the user purchased a ``necklace'' by ``Hello Kitty''. It is reasonable for our method to recommend ``watch'' from the same ``Hello Kitty'' brand. Similar inferences can also be drawn for ``business bag''. Moreover, the interaction with another user and the ``rain'' feature leads to ``umbrella'' being recommended.
In these cases, our method is capable of producing relevant recommendations along with the explainable paths via explicit KG reasoning.

\section{Related Work}\label{sec:related}
There are two main research lines related to our work: KG-based explainable recommendation and multi-behavior recommendation. 

\vspace{0.5em}
\noindent\textbf{KG-based Explainable Recommendation}\quad
Explainable recommendation \cite{zhang2018explainable,zhang2014explicit,li2020generating,chen2019generate,chen2019dynamic,chen2019personalized,xian2020neural} refers to a decision-making system that not only provides accurate recommendation results, but also generates explanations to clarify why the items are recommended. 

One line of research focuses on the KG embedding approach.
Several works integrate KG representation learning into the recommendation model~\cite{zhang2016collaborativekdd,Huang2019ExplainableIU,Wang2018DKNDK,huang2018improving,he2020mining}.
Typically, they assume that the recommended items and their attributes can be mapped into a latent vector space along with transnational relations between the them. 
For example, \citet{zhang2016collaborativekdd} propose the CKE model, which incorporates diverse item types information into Collaborative Filtering. 
\citet{Huang2019ExplainableIU} integrate a KG in multimodal formats capturing dynamic user preferences by modeling the sequential interactions over the KG.
\citet{Wang2018DKNDK} consider both semantics and knowledge representations of news contents for improved news recommendation.
\citet{huang2018improving} leverage KG to enhance item representations and
\citet{he2020mining} jointly conduct KG completion and item recommendations.
These methods demonstrate the effectiveness of incorporating KG embedding into recommendation. However, they fail to directly leverage the KG structure to generate reasoning paths as explanations for the recommendation \cite{zhang2018explainable}.

Another line of work explores incorporating KG reasoning into the process of recommendation. 
The graph structure empowers the system to exploit informative features and also to deliver intuitive path-based explanations.
Early works \cite{catherine2017explainable} propose to model logic rules to conduct explicit reasoning over a KG for explainability.
However, the rules are handcrafted and can hardly generalize to unexplored entity correlations.
In contrast, recent approaches adopt deep neural networks to learn a direct mapping among users, items, and other relations in a KG to enable reasoning for explainable recommendation.
Some approaches \cite{ma2019jointly,wang2018ripplenet,wang2019kgat} only use item-side knowledge but neglect the diverse historical activities of users,
while others \cite{cao2019unifying,ai2018learning} isolate path generation and recommendations, so the resulting path may be irrelevant to the actual decision making process.
We argue that both of these two types of methods fail to model user behaviors to conduct an explicit path reasoning process, which makes the recommendation process less intuitive.
Recently, \citet{xian2019kgrl} and \citet{zhao2020leveraging} perform explicit KG reasoning for explainable recommendation via reinforcement learning. 
Although their paths are generated together with the recommended items,
the recommendation performance is limited by the large search space of the KG and the weak guidance of sparse rewards.
In this work, we follow the setting of KG reasoning for explainable recommendation \cite{xian2019kgrl}, but aim to provide better guidance from user history behavior, as confirmed in our experiments.

\vspace{0.5em}
\noindent\textbf{Multi-Behavior Recommendation}\quad
On modern e-commerce platforms, users can interact with the system in multiple forms \cite{Lo2016UnderstandingBT,liao2018tscset,Singh2008RelationalLV,KrohnGrimberghe2012MultirelationalMF}. 
\citet{Lo2016UnderstandingBT} provide several case studies covering the influence of clicking and saving behavior analysis on the final purchase decision.
Existing methods for multi-behavior recommendations may be divided into two categories: collective matrix factorization based approaches and approaches based on learning from implicit interactions. 
\citet{Singh2008RelationalLV} propose factorizing multiple user--item interaction matrices as a collective matrix factorization model with shared item-side embeddings across matrices.
\citet{Zhao2015ImprovingUT} learn different embedding vectors for different behavior types in an online social network.
\citet{KrohnGrimberghe2012MultirelationalMF} share the user embeddings in recommendation based social network data based on the CMF method.
In contrast,
\citet{Loni2016BayesianPR} proposed an extension of Bayesian Personalized Ranking \cite{rendle2009bpr} as multi-channel BPR, to adapt the sampling rule from different types of behavior in the training of standard BPR.
\citet{Guo2017ResolvingDS} proposed sampling unobserved items as positive items based on item--item similarity, which is calculated using multiple types of feedback.
However, none of these methods consider the reasoning framework to provide explainable recommendations, let alone explicitly model diverse user behaviors over KGs on e-commerce platforms.

\section{Conclusion} \label{sec:conclusions}
In this paper, we propose a new coarse-to-fine KG reasoning approach called \approach{} for explainable recommendation.
Unlike traditional KG based recommendations, our method is characterized by first composing a user profile to capture prominent user behaviors in the coarse stage, and then in the fine stage, conducting path reasoning under the guidance of the user profile.
Since the recommendation and path reasoning processes are closely coupled with each other,  the output paths can be regarded as the explanation to the recommendations.
We extensively evaluate our model on several real-world datasets and show that the proposed approach delivers superior results in recommendation performance.
Our code is available from \url{https://github.com/orcax/CAFE}.

\section*{Acknowledgement}
The authors thank the reviewers for the valuable comments and constructive suggestions. This work was supported in part by NSF IIS-1910154. Any opinions, findings, conclusions or recommendations expressed in this material are those of the authors and do not necessarily reflect those of the sponsors.

\bibliographystyle{ACM-Reference-Format}
\bibliography{paper}

\end{document}